# Catching up with Method and Process Practice: An Industry-Informed Baseline for Researchers

Jil Klünder[1], Regina Hebig[2], Paolo Tell[3], Marco Kuhrmann[4], Joyce Nakatumba-Nabende[5], Rogardt Heldal[6], Stephan Krusche[7], Masud Fazal-Baqaie[8], Michael Felderer[9], Marcela Fabiana Genero Bocco[10], Steffen Küpper[11], Sherlock A. Licorish[12], Gustavo Lòpez[13], Fergal McCaffery[14], Özden Özcan Top[15], Christian R. Prause[16], Rafael Prikladnicki[17], Eray Tüzün[18], Dietmar Pfahl[19], Kurt Schneider[20] and Stephen G. MacDonell[21]

[1&20]Leibniz University Hannover, Germany, Email: {jil.kluender, kurt.schneider}@inf.uni-hannover.de; [2&6]Chalmers | University of Gothenburg, Sweden, Email: {regina.hebig, heldal}@cse.gu.se, [3]IT University Copenhagen, Denmark, Email: pate@itu.dk, [4&11]Clausthal University of Technology, Germany, Email: {marco.kuhrmann, steffen.kuepper}@tu-clausthal.de, [5]Makerere University, Uganda, Email: jnakatumba@cis.mak.ac.ug, [7]Technical University of Munich, Germany, Email: krusche@in.tum.de, [8]Fraunhofer IEM, Germany, Email: masud.fazal-baqaie@iem.fraunhofer.de, [9]University of Innsbruck, Austria, Email: michael.felderer@uibk.ac.at, [10]University of Castilla-La Mancha, Spain, Email: marcela.genero@uclm.es, [12]University of Otago, New Zealand, Email: sherlock.licorish@otago.ac.nz, [13]University of Costa Rica, Costa Rica, Email: gustavo.lopez_h@ucr.ac.cr, [14&15]Dundalk Institute of Technology & Lero, Ireland, Email: {fergal.mccaffery, ozden.ozcantop}@dkit.ie, [16]DLR Space Administration, Germany, Email: christian.prause@dlr.de, [17]Pontifícia Universidade Catòlica do Rio Grande do Sul, Brazil, Email: rafael.prikladnicki@pucrs.br, [18]Bilkent University, Turkey, Email: eraytuzun@cs.bilkent.edu.tr, [19]University of Tartu, Estonia, Email: dietmar.pfahl@ut.ee, [21]Auckland University of Technology, New Zealand, Email: smacdone@aut.ac.nz

**Abstract**

*Software development methods are usually not applied by the book. Companies are under pressure to continuously deploy software products that meet market needs and stakeholders' requests. To implement efficient and effective development processes, companies utilize multiple frameworks, methods and practices, and combine these into hybrid methods. A common combination contains a rich management framework to organize and steer projects complemented with a number of smaller practices providing the development teams with tools to complete their tasks. In this paper, based on 732 data points collected through an international survey, we study the software development process use in practice. Our results show that 76.8% of the companies implement hybrid methods. Company size as well as the strategy in devising and evolving hybrid methods affect the suitability of the chosen process to reach company or project goals. Our findings show that companies that combine planned improvement programs with process evolution can increase their process' suitability by up to 5%.*

**Index Terms:** software development, software process, hybrid methods, survey research

## 1. INTRODUCTION

For decades, software companies, teams, and even individual developers have sought approaches that enable efficient and effective software development. Since the 1970's, numerous processes have been proposed. The community started with the Waterfall model [1], then the Spiral model [2], followed by agile methods and lean development approaches [3]. Since the early 2000s, few innovative software development approaches were proposed, but several proposals for scaling agile methods, e.g., SAFe or LeSS, were published. Meanwhile, an increasing number of studies showing that modern software development is neither purely "traditional" nor "agile" can be found reflecting that companies use processes comprised of various development practices [4], [5].

*Problem Statement:* Research that focuses on agile methods and practices only cannot support practitioners who are faced with the reality of *hybrid development methods*. Similarly, the 100+ tailoring criteria [6], [7] for processes established in the last decade seem to have no relevance for practitioners who are devising hybrid methods and seeking im- mediate and practical solutions to solve short-term problems. Thus, process-related research has lost momentum as it is no longer aligned with the concerns of practice.

*Objective:* In response to the situation above, our objective is *to understand how and why practitioners devise hybrid development methods*. Our goal is to set a new baseline for the next decade of evidence-based research on software development approaches driven by practice.

*Contribution:* Based on an online survey comprising 732 data points we study the use of hybrid methods and the factors influencing the suitability of development approaches for reaching goals. According to our results, 3/4 of the companies use a hybrid method, and company size and strategies to devise hybrid methods influence the suitability of the approach to achieve defined goals.

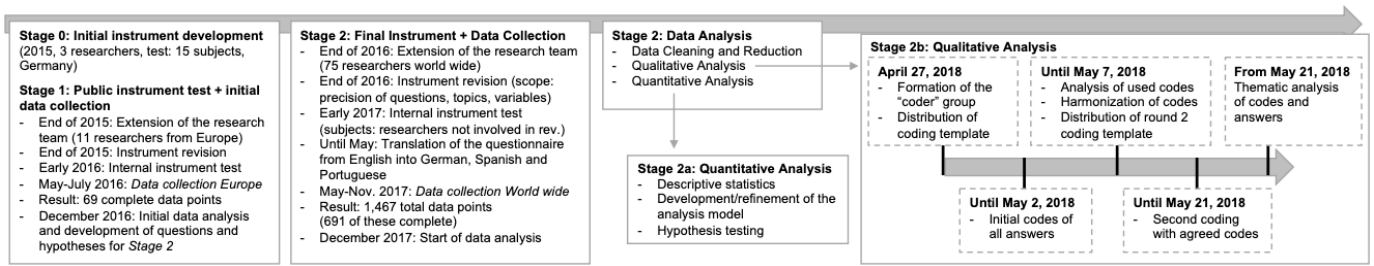

Fig. 1. Overview of the research design.

*Context:* This research is based on the HELENA study[1], which is a large-scale international survey in which 75 researchers and practitioners from 25 countries participated. We give further details on the implementation of the HELENA study in Section 3-B.

*Outline:* The rest of this paper is organized as follows: In Section 2, we present related work. Section 3 presents our research method. In Section 4, we present our results, which are discussed in Section 5. We conclude in Section 6.

## 2. RELATED WORK

The use of software development processes has been studied since the 1970s, when the first ideas to structure software development appeared [1], [2]. Since then, a growing number of approaches emerged, ranging from traditional and rather sequential models, to iterative and agile models. Various combinations are used, forming hybrid methods.

In 2003, Cusumano et al. [8] surveyed 104 projects and found many using and combining different development approaches. In an analysis of 12,000 projects, Jones [9] found that both specific design methods and programming language can lead to successful or troubled project outcome. Neill and Laplante [10] found that approximately 35% of developers used the classical Waterfall model. However, projects also used incremental approaches, even within particular lifecycle phases. In 2014, Tripp and Armstrong [11] investigated the "most popular" agile methods and found XP, Scrum, Dynamic Systems Development Method (DSDM), Crystal, Feature Driven Development (FDD), and Lean development among the top methods used. Only a few studies investigate the development of processes over time. One perspective on the use of agile methods is provided by Dingsøyr et al. [12]. They provided an overview of "a decade of agile software development", and motivate research towards a rigorous theoretical framework of agile software development, specifically, on methods of relevance for industry. Such a perspective is given by the VersionOne and the Swiss Agile surveys [13], [14] that investigate the use of agile methods.

In 2011, West et al. [5] coined the term "Water-Scrum-Fall" to describe the process pattern mostly applied in practice at that time. Recent studies point to a trend towards using such combined processes. Garousi et al. [15] as well as Vijayasarathy and Butler [16] found that "classic" processes like the Waterfall model are increasingly combined with agile methods. Solinski and Petersen [17] found Scrum and XP to be the most commonly adopted methods, with Waterfall/XP, and Scrum/XP as the most common combinations. In 2017, Kuhrmann et al. [4] generalized this concept, defining the term "*hybrid development methods*" as "any combination of agile and traditional (plan-driven or rich) approaches that an organizational unit adopts and customizes to its own context needs" [4]. They also confirmed that numerous development processes are applied and combined with each other.

Available studies thus show a situation in which traditional and agile approaches coexist and form the majority of practically used hybrid methods. In contrast, current literature on software processes and their application in practice leaves researchers and practitioners with an increasing amount of research focusing only on agile methods. Traditional models are vanishing from researchers' focus. They only play a role in process modeling, in domains with special requirements (e.g., regulations and norms), or in discussions why certain companies do not use agile methods (cf. [11], [18]).

Empirical data about general software process use, trends in global regions, and detailed information about the combination of processes is missing. To correctly portray the state of practice, empirical data from industry is needed. This paper fills this gap by providing a big picture of the use of hybrid methods with respect to various development contexts (industry sector, domain, company size) and different constraints companies face.

## 3. RESEARCH METHOD

We describe the overall research design following the steps shown in Fig. 1 by presenting the research objective and research questions, followed by a description of the procedures executed for the collection and analysis of data.

### A. Research Objective and Research Questions
Our research objective is to understand why and how practitioners use hybrid methods in practice. For this, we conducted a large-scale international online survey to study

---
[1] HELENA: Hybrid dEveLopmENt Approaches in software systems development, online: https://helenastudy.wordpress.com.

(i) which hybrid methods are practically used, (ii) how practitioners devise such methods, and (iii) which strategies used to devise such methods are beneficial. Emerging from the first stage of our study (Fig. 1), the research questions are:

*RQ1:* *Which software development approaches are used and combined in practice?* This question aims to determine the state of practice to lay the foundation for our research. Specifically, we study which methods, frameworks and practices are used in practice and if they are combined.

*RQ2:* *Which strategies are used to devise hybrid methods in practice?* This question aims at investigating why and how hybrid methods are defined in practice, i.e., if specific combinations are developed intentionally, if they evolve over time, or if they were devised in response to specific situations. Furthermore, we examine which goals are addressed by the chosen development approach.

*RQ3:* *Are there differences between the strategies used to devise hybrid methods regarding gained benefits?* When a hybrid method is devised, this happens in response to an implicit or explicit purpose, e.g., a need to improve communication. This research question aims to analyze whether strategies to devise hybrid development methods are comparable with regard to gained benefits, i.e., that they equally allow practitioners to devise a method that can fulfil the targeted purpose.

### B. Instrument Development and Data Collection

We used the survey method [19] to collect our data. We designed an online questionnaire to solicit data from practitioners about the processes they use in their projects. The *unit of analysis* was either a project or a software product.

*1) Instrument Development:* We used a multi-staged approach to develop the survey instrument. Initially, three re- searchers developed the questionnaire and tested it with 15 German practitioners to evaluate suitability (Fig. 1, Stage 0). Based on the feedback, a research team of eleven researchers from across Europe revised the questionnaire. A public test of the revised questionnaire, that included up to 25 questions, was conducted in 2016 in Europe (Fig. 1, Stage 1). This public test yielded 69 data points, which were analyzed and used to initiate the next stage of the study [4]. In Stage 2, the research team was extended, with 75 researchers from all over the world now included. The revision of the questionnaire for Stage 2 was concerned with improving structure and scope, e.g., relevance and precision of the questions, value ranges for variables, and relevance of the topics included. The revised questionnaire was translated and made available in English, German, Spanish, and Portuguese. Further details of the instrument are presented in [20].

*2) Instrument Structure:* The final questionnaire consisted of five parts (with number of questions): Demographics (10), Process Use (13), Process Use and Standards (5), Experiences (2) and Closing (8). In total, the questionnaire consisted of up to 38 questions, depending on previously given answers [20].

*3) Data Collection:* The data collection period was May to November 2017 following a *convenience sampling strategy*

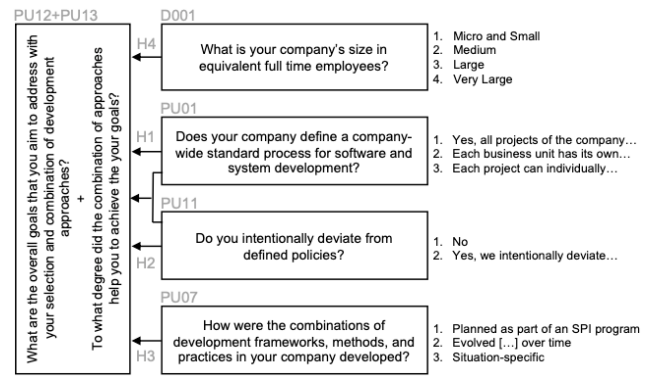

Fig. 2. Analysis model for quantitative analysis. The model shows the six questions (incl. question IDs), the value ranges and the linked hypotheses.

[19]. The survey was promoted through personal contacts of the 75 participating researchers, through posters at conferences, and through posts to mailing lists, social media channels (Twitter, Xing, LinkedIn), professional networks and websites (ResearchGate and researchers' (institution) home pages).

### C. Data Analysis

As illustrated in Fig. 1, the data analysis approach applied to the survey results included three main steps, which we present in the following subsections.

*1) Data Cleaning and Data Reduction:* In total, the survey yielded 1,467 answers, and 691 participants completed the questionnaire. Hence, as a first step, we analyzed the two datasets (partially and completely answered questionnaires) and performed different analyses (descriptive statistics, two researchers) to investigate the effects of using the partial or the complete dataset. In the second step, two researchers reviewed the data again in the context of the research questions and used the results to develop a *suggested* dataset, which adds elements from the partial dataset to the dataset containing the complete answers. Finally, from the 1,467 answers, we selected 732 answers (49.9%) for inclusion in our data analysis. Each answer forms a data point that consists of 206 variables (plus meta data). The complete data set can be found in [20].

*2) Quantitative Analysis:* The quantitative analysis employed several instruments, e.g., descriptive statistics and hypothesis testing. We summarize these instruments and we describe how we handled the data to support these instruments.

**a) Data Handling and Data Aggregation:** At first, we cleaned, aggregated and analyzed the data. Specifically, we analyzed the data for *NA* and *-9* values. While *NA* values indicate that participants did not provide information for optional questions, *-9* values indicate that participants skipped a question. Depending on the actual question, *-9* values were transformed into *NA* values, or the data point was excluded from further analyses as we considered the question incompletely answered. For example, if the question about the *goals addressed by a combination of methods* (Fig. 2, PU12) was answered, but the follow-up question for the *suitability of the combination regarding the goals set* (Fig. 2, PU13) was not, this data point was discarded. Furthermore, in the question on company size (Fig. 2, D001), we integrated the category *Micro* with the category *Small*, which results in a new category *Micro and Small (1–50 employees)*.

TABLE I. Hypotheses To Test The Suitability Of Hybrid Methods (RQ3).

| Hypotheses | |
|---|---|
| $H1_0$ | The suitability of a chosen development approach does not depend on having a company-wide process. |
| $H2_0$ | The suitability of a chosen development approach does not depend on deviating from defined policies. |
| $H3_0$ | The suitability of a chosen development approach does not depend on the evolution of the combination. |
| $H4_0$ | The suitability of a chosen development approach does not depend on the company size. |

**b) Development/Refinement of the Analysis Model**: Figure 2 shows the *analysis model* consisting of six questions in the questionnaire, which we developed to provide a framework for the (non-descriptive) quantitative analysis. In the rest of the paper, we use short versions of the questions from Fig. 2 (together with the question ID to allow for a mapping). The center of the analysis model (Fig. 2, left) is the combination of the two questions PU12 and PU13 asking about the goals set by combining development approaches in a specific way and the suitability of this combination. The remaining four questions were selected to study influence factors and dependencies, e.g., does the company size (D001) or a specific way of devising a hybrid method (PU07) influence the suitability. The actual data analysis using our model was carried out in two steps: (i) we explored the data on a per-question basis, i.e., variables were analyzed in an isolated manner, and (ii) we paired the different questions.

**c) Hypothesis Testing:** The final step in the quantitative analysis (Fig. 1) was the hypothesis testing. Table I summarizes the hypotheses tested in the context of RQ3. To test the hypotheses, we analyzed the data with statistical tests chosen based on certain pre-conditions. Before the actual test, we tested each variable for normality with the Shapiro-Wilk test[2]. To test the hypotheses H1 to H4 we had to determine the *suitability* of the respective hybrid method in relation to the goals participants targeted while devising it. Participants could choose from 18 goals, and for each selected goal g, participants rated the suitability of the actual hybrid method on a 10-point scale: $suit_g \in \{1, \ldots, 10\}, 1 \leq g \leq 18$. Since participants could select a different number of goals, the suitability per participant p was standardized to abstract from the number of individually selected goals: $suit_g(p) \in [0,1]$. To apply the analysis model, we calculated the total suitability for a given participant and the overall suitability of a goal:

$$\text{suit}_{\text{total\_participant}}(p) = 0.1 \cdot \underset{g}{\text{avg}}\{\text{suit}_g(p,g)\}$$
$$\text{suit}_{\text{total\_goal}}(g) = 0.1 \cdot \underset{p}{\text{avg}}\{\text{suit}_g(p,g)\}$$

All variables of the analysis model (Fig. 2; PU01: company-wide policies, PU11: deviation from these policies, PU07: permutations of the different strategies to devise a hybrid method, and D001: company size) were individually tested against the suitability calculated for the different groups. For this, we categorized the data and tested the respective means of the suitability for significant differences on a per-variable basis using Pearson's $\chi^2$ test[3] and the Kruskal-Wallis test[4]. Finally, we tested combinations of the variables using the Kruskal-Wallis test. If evidence to reject the null-hypotheses was found, effect sizes were calculated using $\varepsilon^2$ as suggested by Tomczak and Tomczak [21]. For interpretation we apply commonly used thresholds, inspired by Cohen's interpretation of Pearson's *r* [22] and adapted the character of $\varepsilon^2$: an effect size of $0.01 \leq \varepsilon^2 < 0.08$ is considered small, $0.08 \leq \varepsilon^2 < 0.26$ is considered medium, and $0.26 \leq \varepsilon^2$ is considered large.

*3) Qualitative Analysis:* In the analysis it became clear that deviating from defined policies might not lead to as much benefit as other strategies. However, as deviation was reported in many cases, we decided to conduct additional qualitative analyses focusing on the reasons why developers intentionally deviate from policies (optional free-text comment to PU11). In addition, we investigated the free-text answers for reasons to devise hybrid methods (PU06). Both analyses were performed on the complete data set with 731 data points (one data point was discarded due to missing answers).

The qualitative analysis was challenging due to the large number of data points (267 out of 731 participants provided answers for PU11 and 89 for PU06) as well as the language diversity among the answers received (English, German, Spanish, and Portuguese). We addressed this by distributing the analysis activity across a core team of three researchers and an extended team of 12 additional researchers who focused on coding the data. Together, we performed an analysis based on coding, following the process shown in Fig. 1. The coding process included five steps: (i) A core team of three researchers prepared a coding template and organized the coding (taking language skills/preferences into account) and the distribution of the data such that two independent codings per data point were performed. (ii) All 15 coders conducted the coding. In total, this step yielded 123 codes for PU11 and 59 codes for PU06—all codes in English. (iii) The core group analyzed the codes and provided a harmonized set of 56 codes for PU11 and 50 codes for PU06 to the coding group. (iv) The coders re-coded the data using the agreed codes. (v) The core group performed a thematic analysis on the coded data. In total, nine themes of reasons for deviation were named for PU11, and 38 additional reasons for devising hybrid methods were found for PU06, including 16 reasons mentioned by more than one participant.

## 4. RESULTS

After a characterization of the study population, we present the results organized according to the research questions.

---

[2] The Shapiro-Wilk test is used to test if a sample comes from a normally distributed population (null hypothesis).

[3] Pearson's $\chi^2$ tests whether two variables are independent (null hypothesis).

[4] The Kruskal-Wallis test is a non-parametrized test that can be applied for comparing more than two groups. The test investigates if there are no differences between the groups (null hypothesis).

Table II. Overview Of Company Size And Participants' Roles (N=732).

| | Micro/Small | Medium | Large | Very Large | no Info | Σ | % |
|---|---|---|---|---|---|---|---|
| Developer | 45 | 49 | 54 | 47 | 1 | 196 | 26.8 |
| Project/Team Manager | 32 | 42 | 33 | 36 | – | 143 | 19.5 |
| Product Manager/Owner | 24 | 13 | 14 | 18 | – | 69 | 9.4 |
| Architect | 15 | 15 | 19 | 14 | – | 63 | 8.6 |
| *Other* | 7 | 17 | 22 | 17 | – | 63 | 8.6 |
| C-level Management (e.g., CIO, CTO) | 26 | 12 | 8 | 4 | – | 50 | 6.8 |
| Scrum Master/Agile Coach | 10 | 10 | 8 | 21 | – | 49 | 6.7 |
| Analyst/Requirements Engineer | 12 | 11 | 11 | 4 | 2 | 40 | 5.5 |
| Quality Manager | 5 | 5 | 19 | 7 | – | 36 | 4.9 |
| Tester | – | 6 | 7 | 1 | – | 14 | 1.9 |
| Trainer | 4 | 2 | 3 | – | – | 9 | 1.2 |
| Σ | 180 | 182 | 198 | 169 | 3 | 732 | |
| % | 24.6 | 24.9 | 27.0 | 23.1 | 0.4 | | 100 |

### A. Study Population and Descriptive Statistics

As described in Section 3-C1, 732 data points were included in the data analysis. Answers were included from 46 countries, with 19 countries providing 10 or more answers and 13 countries providing 20 or more answers. Most answers were received from Germany (127), Brazil (80), Argentina (65), Costa Rica (51), and Spain (50). The average time (median) for completing the questionnaire was 23:36 minutes.

*1) Respondent Profiles:* Table II provides an overview of the participants. The largest groups are developers (26.8%) and project/team managers (19.5%). The 63 participants who selected "other" as their role described themselves as functional safety manager, data scientist, DevOps engineer, or having multiple roles. Table II also shows the distribution of the participants across the different company sizes, showing companies of all sizes equally present in the result set. Three participants did not provide information about the company size. Additionally, 59.8% of the participants have 10 or more years of experience in software and system development and only 7.8% have two years or less of experience.

*2) Project and Product Profiles:* The unit of analysis in the study at hand was a specific project or product. In total, 60.2% of the participants classified their project as having one person year or more of effort. Regarding the target domain, web applications and services (26.8%) as well as financial services (24.0%) are the most frequently mentioned. The remaining tar- get domains include mobile applications (16.4%), automotive software (10.4%), logistics (7.2%), and space systems (4.6%). The 11.9% in the category "other target domains" named, among others, agriculture, industry/production automation, human resources, and stores/retail.

### B. RQ1: Software Development Approaches

We are interested in capturing the state of practice in the use of development frameworks, methods and practices, and in analyzing whether these are combined with each other. Of the 732 participants, 562 (76.8%) stated that they combine different development approaches into a hybrid method. Two questions asked about the use of 24 development frameworks and methods, and 36 development practices, respectively. Participants stated whether they know the frameworks, methods and practices as well as if they use a framework or practice, and to what extent. An overview of the knowledge and use of frameworks, methods, and practices in hybrid methods is shown in Fig. 3 (frameworks and methods) and Fig. 4 (practices). For both figures, we only consider answers of the 562 cases that reported using hybrid methods.

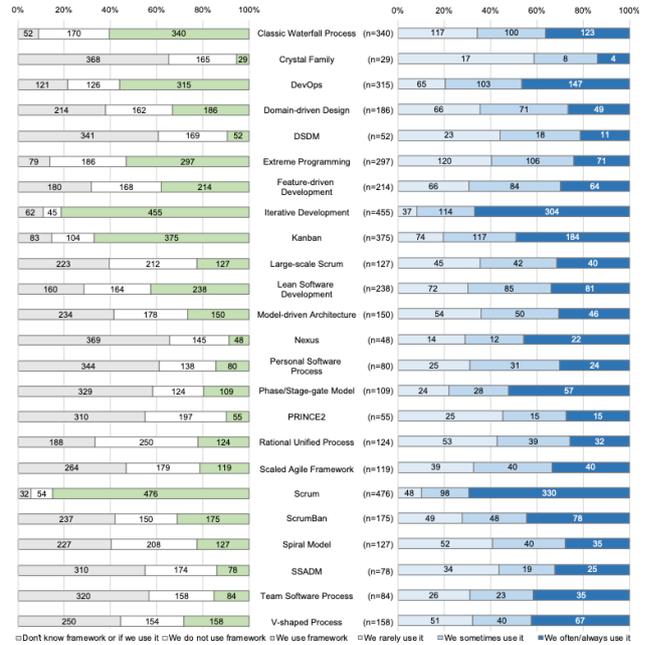

Fig. 3. Overview of the knowledge and usage of frameworks and methods in hybrid methods. The left part of the figure shows the breakdown for knowledge and usage. The right part breaks down the "We use framework"-statements into the usage frequency for the individual frameworks/methods.

> **Finding 1:** In total, 76.8% (562 out of 732) of the reported cases state to use hybrid development methods.

Table III shows that Scrum, Iterative Development, Kanban, the "classic" Waterfall model and DevOps are the most frequently mentioned development methods or frameworks within hybrid methods and also within the whole data set (including non-hybrid development methods; [20]). Furthermore,

Table III provides the rank in the category *"We often/always use"* (column "% use"), which reads as follows: of the 84.7% of participants stating that they use Scrum, 69.3% often or always use Scrum. Each of the six frameworks and methods in Table III is used by at least 50% of the participants reporting to use hybrid methods. At the other end of the spectrum, PRINCE2 (9.7%), DSDM (9.2%), and Nexus (8.5%), and the Crystal Family (5.1%) received the smallest number of mentions. Notable, the numbers do not change much when considering the whole data set as also shown in Table III.

The practices (Fig. 4) draw a more diverse picture. Of the 36 practices provided in the questionnaire, 28 are used by more than 50% of the participants who use hybrid methods. The least used practices are Automated Theorem Proving (14.9%), Model Checking (25.6%) and Formal Estimation (33.8%). The most frequently mentioned development practices are Code Reviews (94.5%) and Coding Standards (93.4%), followed by Release Planning (89.3%),

Table III. Most Frequently Mentioned Development Frameworks and Methods Within Hybrid Methods (Threshold 50%) and Share Of Cases That Reported to Often or Always Use That Framework (Respective Data For The Whole Dataset In Parentheses).

| Framework | Rank | % Use | % Often/always used when used |
|---|---|---|---|
| Scrum | 1 (1) | 84.7 (81.6) | 69.3 (71.7) |
| Iterative Development | 2 (2) | 80.9 (76.1) | 66.8 (65.9) |
| Kanban | 3 (3) | 66.7 (63.9) | 49.1 (49.6) |
| Classic Waterfall | 4 (4) | 60.5 (55.2) | 36.2 (36.4) |
| DevOps | 5 (5) | 56.0 (54.4) | 46.7 (48.2) |
| Extreme Programming | 6 (6) | 52.8 (50.3) | 23.9 (24.2) |

Table IV. Company-Wide Policies (Pu01) And Deviation (PU11).

| Framework | Rank | % Use | % Often/always used when used |
|---|---|---|---|
| Scrum | 1 (1) | 84.7 (81.6) | 69.3 (71.7) |
| Iterative Development | 2 (2) | 80.9 (76.1) | 66.8 (65.9) |
| Kanban | 3 (3) | 66.7 (63.9) | 49.1 (49.6) |
| Classic Waterfall | 4 (4) | 60.5 (55.2) | 36.2 (36.4) |
| DevOps | 5 (5) | 56.0 (54.4) | 46.7 (48.2) |
| Extreme Programming | 6 (6) | 52.8 (50.3) | 23.9 (24.2) |

Prototyping (88.8%), Automated Unit Testing (86.6%), and Refactoring (85.7%). Summarizing, companies frequently use a variety of practices and, with a few exceptions, most of the practices are widely used.

> **Finding 2:** Companies combine frameworks, methods and practices to form hybrid methods. Scrum, Iterative Development, Kanban, Waterfall and DevOps are the most frequently used frameworks and methods.

### C. RQ2: Strategies to Devise Hybrid Methods

In this section, we study why hybrid methods are used and how they are devised using the analysis model from Fig. 2.

*1) Policies and Deviation:* First, we studied whether companies have standard processes or policies defined (PU01) and if the participants intentionally deviate from such policies (PU11). Table IV shows that about half of the companies have a company-wide standard process (45.8%), 19% of the participants have standard processes defined at the business unit level, and approx. 1/3 of the participants (35.2%) individually decide which process to follow. Yet, only 37.4% of the participants state that they intentionally deviate from their defined policies.

*2) Motivation for Devising Hybrid Methods:* Approximately 3/4 of the participants devise hybrid methods to run their projects. Hence, we study reasons for devising such methods. In the questionnaire, participants were asked two key questions (Fig. 2; PU12 and PU13). Question PU12 provided participants with 18 pre-defined goals (cf. Table VI) for which they could state if these goals are drivers for the chosen hybrid method. To ensure they did not miss a goal, PU12 was complemented with an optional free-text field to collect further goals. For each goal selected in question PU12, participants were presented their selection in PU13 for the purpose of evaluating if a specific goal was achieved through the hybrid method (the analysis of the hybrid methods' suitability is presented in Section 4-C4).

In a nutshell, independent of whether or not respondents deviated from defined company policies, the most frequently named goals are: *improved productivity*,

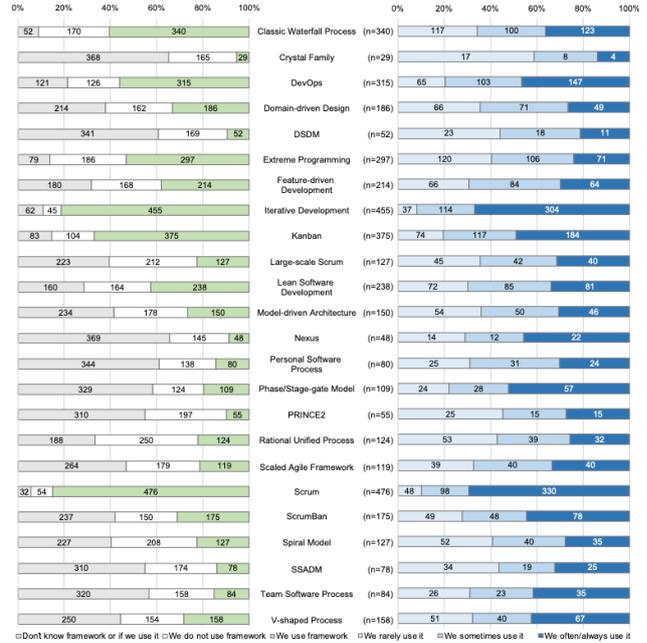

Fig. 4. Overview of the knowledge and usage of development practices in hybrid methods. The left part of the figure shows the breakdown for knowledge and usage. The right part breaks down the "We use practice"-statements into the usage frequency for the individual practices.

*improved external product quality*, *improved planning and estimation*, *improved frequency of delivery to the customer* and *improved adapt-ability and flexibility of the process to react to change*. The additional open question revealed some extra goals of which, however, none represents a relevant addition.

> **Finding 3:** The most popular goals addressed by companies are improved productivity, improved external product quality, and improved planning and estimation.

*3) Strategies to Devise Hybrid Methods:* Several strategies are used to devise hybrid methods. Table V shows that 37.8% of the participants developed their hybrid method through a choreographed software process improvement (SPI) initiative.

However, most hybrid methods evolve over time based on learning from past projects (78.5%).

> **Finding 4:** The most common way to devise hybrid development methods is evolution, followed by planning as part of SPI initiatives.

*4) Suitability of Devised Hybrid Methods:* Table VI shows the 18 pre-defined goals from which the participants could choose (Fig. 2, PU12) and the suitability of the devised hybrid methods for cases with an intentional deviation from a defined company policy. The overall suitability is the average of all suittotal goal(g) and results in 0.6575, i.e., for all cases, the hybrid method was suitable to achieve the goals set to approx. 66%. Participants that do not deviate from a company policy tend to perceive their hybrid methods slightly more suitable to achieve their goals (67.47%, SD=16.48%, n=303) than those who deviate (63.55%, SD=18.30%, n=228).

> **Finding 5:** Hybrid methods devised by practitioners are suitable to achieve their goals with a probability of approx. 66%.

TABLE V. Strategies To Devise (Evolution Of) Hybrid Methods (PU07).

| Question/Answer | Quantities | |
|---|---|---|
| *How were the combinations of development frameworks, methods, and practices in your company developed? (PU07, n=543)* | | |
| Planned as part of a SPI program | 205 | 37.8% |
| Evolved as learning from past projects over time | 426 | 78.5% |
| Situation-specific | 46 | 8.5% |

Table VI. Overview of The Set Goals and Suitability of Hybrid Methods to Achieve Them (N: Number Of Participants That Have Set This Goal; Med, Mean, SD: The Suitability in %).

| Goal: *Improved...* | n | Med | Mean | SD |
|---|---|---|---|---|
| Productivity | 165 | 70 | 61.2 | 24.03 |
| External product quality | 146 | 70 | 68.2 | 20.07 |
| Planning and estimation | 144 | 60 | 59.3 | 21.99 |
| Frequency of delivery to customers | 143 | 70 | 69.9 | 23.48 |
| Adaptability and flexibility of the process to react to change | 131 | 80 | 73.1 | 21.49 |
| Time to market | 122 | 70 | 64.9 | 25.30 |
| Client involvement | 118 | 70 | 70.7 | 20.74 |
| Internal artifact quality | 112 | 70 | 67.1 | 21.42 |
| Project monitoring and controlling | 110 | 70 | 64.1 | 21.94 |
| Knowledge transfer and learning | 105 | 70 | 63.4 | 22.53 |
| Employee satisfaction | 102 | 70 | 67.3 | 23.68 |
| Risk management | 83 | 60 | 58.4 | 20.98 |
| Reuse for project artifacts | 79 | 60 | 59.8 | 20.00 |
| Return-on-investment cycles | 77 | 60 | 60.9 | 22.49 |
| Maturity of the company | 61 | 60 | 58.4 | 24.85 |
| Staff education and development | 59 | 70 | 62.4 | 24.09 |
| Tool support | 50 | 60 | 56.8 | 16.59 |
| Ability of the company to develop critical systems | 40 | 60 | 59.8 | 26.84 |

*5) Reasons for Deviation:* Our data suggests that deviating from a defined company policy is disadvantageous even though deviations are reported in many cases. To study reasons for deviations, we conducted a thematic analysis on the optional free-text answers that complement the question PU11. Following the coding procedure (Section 3-C3), we identified nine themes (a threshold was set to five instances of a code). From the 267 data points, 65 have been assigned to multiple codes and, thus, have been assigned to multiple themes.

In total, 83 participants stated they have *explicit goals* for deviating from the policies. Such goals often go hand in hand with the motivation for creating hybrid methods such as avoiding overhead (20) or compensating time pressure (18) and resource constraints (7):

> "We make appearances of following process and procedures, but really just try to do what we can based on resource and skills constraints, and because processes are defined by committees who don't actually have to deliver. This is the "How I have survived in this game for 30 years" answer." [participant 1453]

Factors like flexibility (17), costs-benefit/efficiency (9), quality (8), and speed (6) were also mentioned.

***Accommodating context factors*** were stated by 75 participants. Such factors include project specific factors (45), different or new technologies, domains, or tools (13), and different teams (11). *Accommodating the client, market or business* as a reason to deviate was mentioned by 72 participants, notably due to partner/client requests (32). In 38 cases, participants named *shortcomings of the company's standard process* as a reason to deviate as, for instance, the standard process is described too abstract for direct implementation:

> "Processes used case by case, the process is defined for the largest possible project and everything is not applicable for smaller changes and projects. Then selected parts can be removed or done differently." [participant 2632]

Another nine participants stated that the standard process was outdated or inappropriate, and five participants stated that there is no standard process at all. *Process improvement* was mentioned by 16 participants, e.g., for implementing a continuous improvement approach and to build on experience. Closely related, *experimentation* as a reason to deviate was mentioned by 14 participants with the purpose of trying new processes and methods. Another driver for deviation reported by 12 participants is to *create a fit between different processes, organizations, or tools*, whereas the need to align project processes and client processes was highlighted:

> "Our own processes and those of the customer had to be reconciled." [participant 2960; translated from German]

Some reasons were named by only single participants, such as organizational politics or deviation by mistake.

> **Finding 6:** Deviation for a process is most commonly motivated by explicit goals, context factors, the need to accommodate the client, market, or business, and issues with the standard process of the company.

### D. RQ3: Differences in Strategies to Devise Hybrid Methods

Companies use different development approaches in combination as hybrid methods (Section 4-B) and use different strategies to devise them (Table V). We analyze these strategies with respect to their potential influence on the suitability of using hybrid methods to achieve certain goals.

*a) Isolated Test of Variables:* Using our analysis model (Section 3-C2b), we explore the data by studying the variables from Fig. 2, i.e., company size, company-wide policies, deviation from such policies and the way of devising hybrid methods. For participants (n=562) stating that they use different processes in combination, we test the four variables in relation to the suitability suitg of reaching the goals set through the use of hybrid methods. As described in Section 3-C2c, for the variables of interest, we tested the normality of the distribution of the suitability suitg with the Shapiro-Wilk test (W = 0.95714, p-value = $2.559 \times 10^{-11}$) and concluded that the non-parametrized Kruskal-Wallis test should be used for further analyses. Table VII summarizes the results of the tests for the isolated influence of the company-wide process (H1), the deviation (H2), the evolution (H3) and the company size (H4) on the suitability of the process. Table VII also shows that only the intentional deviation (H2) and company size (H4) show significant differences in the treatment groups— both show small, but non-negligible effect sizes. Table VIII shows the average suitability to reach the goals in dependence of the company size and the intentional deviation.

> **Finding 7:** For projects in small companies, we found an 3.2% increased chance to reach the goal compared to medium companies and 6.5% compared to large companies.

**Finding 8:** For projects that intentionally deviate from the process, we found an 3.9% decreased chance to reach the goals compared to projects that do not intentionally deviate.

*b) Combined Analysis of Variables:* After the isolated exploration of the variables of interest, we studied the combi- nation of variables, i.e., are there effects on the suitability of devising a hybrid method for companies that, for instance, have a company-wide policy defined from which participants intentionally deviate. Furthermore, we analyzed specific (combined) strategies to devise a hybrid method. To test the potential effects of deviations from defined policies (Fig. 2; PU01, PU11), we first confirmed with the Kruskal-Wallis test that the combination of PU01 and PU11 is not significant ($\chi^2 = 9.481$, df = 5, p-value = 0.09135).

Since the question for the evolution of the company's development approach is a multiple-choice one we built the permutations and compared the groups with each other. Given the differences in the samples, we decided to focus on the three largest groups (Table IX), which were extracted and compared to the rest of the sample (Table X). Among the groups, the group [1,1,0], i.e., participants who devise their hybrid method in a planned and evolutionary manner driven by experience gathered in past projects, showed a significant difference with a small, but non-negligible effect size ($\varepsilon^2$=0.0127). The other two groups did not show significant results.

**Finding 9:** Projects that devise hybrid processes applying both strategies (planning as part of SPI and evolving it based on experience) have an approx. 5% better chance to reach the goals set for devising the process.

## 5. DISCUSSION

We discuss our findings, research questions, threats to validity and future directions of research.

### A. Answering the Research Questions

To understand how practitioners use hybrid methods in practice, we formulated three research questions (Section 3-A). Based on our findings, we answer these as follows:

*RQ1:* Combining different development frameworks, methods and practices is the state of practice. Almost all methods and practices are used to form hybrid methods.

TABLE VII. Results Of The Kruskal-Wallis Test For H1 To H4.

| Id | Results | Decision |
|---|---|---|
| $H1_0$ | $\chi^2 = 2.78$, df = 2, $p = 0.2491$ | no statement |
| $H2_0$ | $\chi^2 = 5.5692$, df = 1, $p = 0.01828$, $\varepsilon^2$=0.013 | reject |
| $H3_0$ | $\chi^2 = 10.93$, df = 6, $p = 0.09057$ | no statement |
| $H4_0$ | $\chi^2 = 18.83$, df = 4, $p = 0.0008487$, $\varepsilon^2$=0.0355 | reject |

Table VIII. Average Suitability By Company Size (H4) and Deviation (H2).

| Variable | Value | n | $suit_g$ in % |
|---|---|---|---|
| Company Size | Micro and Small | 127 | 70.19734 |
| | Medium | 131 | 66.91896 |
| | Large | 149 | 63.59917 |
| | Very Large | 124 | 62.71493 |
| | NA | 1 | 46.66667 |
| Intentional Deviation | No | 303 | 67.46700 |
| | Yes | 228 | 63.54623 |

TABLE IX. Average Suitability by Strategy to Devise, I.E. The Evolution of a Hybrid Method (PU07).

| Variable | Permutations | n | $suit_g$ in % | |
|---|---|---|---|---|
| Strategies to devise[a] | [0,0,0] | 0 | 0 | |
| | [0,0,1] | 31 | 58.19790 | |
| | **[0,1,0]** | 294 | 65.65112 | (select) |
| | [0,1,1] | 8 | 65.94692 | |
| | **[1,0,0]** | 81 | 63.76353 | (select) |
| | [1,0,1] | 3 | 52.72727 | |
| | **[1,1,0]** | 111 | 70.06822 | (select) |
| | [1,1,1] | 4 | 61.82540 | |

[a] The answers to question PU07 (Fig. 2) represent the permutations of the multiple choice answer options [planned, evolved, situation-specific] with 1=*selected* and 0=*not selected*. For instance, [1,0,0] includes all participants who *only* use a planned SPI-approach to devise a hybrid method.

Table X. Suitability of Addressed Goals Based on The Different Strategies to Devise (Evolution Of) a Hybrid Method (PU07).

| | [1,1,0] | | [0,1,0] | | [1,0,0] | |
|---|---|---|---|---|---|---|
| | n | $suit_g$ | n | $suit_g$ | n | $suit_g$ |
| Group | 111 | 0.7006822 | 294 | 0.6565112 | 81 | 0.6376353 |
| Rest | 421 | 0.6461631 | 238 | 0.6588072 | 451 | 0.6611130 |
| $\chi^2$, df | 6.755, 1 | | 0.068546, 1 | | 0.71172, 1 | |
| p-value | 0.009348 | | 0.7935 | | 0.3989 | |

The methods most often used as ingredients in hybrid methods are *Scrum* and *Iterative Development*.

*RQ2:* The most common strategy to devise hybrid methods is to evolve the process based on experience, followed by planning a hybrid method as part of an SPI initiative. Explicitly devising a hybrid method towards a specific project situation is seldom. However, such situations are drivers for process deviation.

*RQ3:* Several strategies are used to devise hybrid methods. The strategies are influenced by a number of context factors. Our data shows that devising hybrid methods through a planned SPI approach including experience from past projects increases the probability of reaching the set goals, i.e., to devise a meaningful method. Our data also shows that deviations from defined policies might decrease the probability of reaching the goals. However, the rather small effect sizes indicate that these results have to be interpreted with care. While they indicate an impact, it is not clear how much potential impact there is when improving the strategies to devise hybrid methods. Hence, future studies should conduct deeper analyses to further investigate this indication and show under what conditions improvement can be reached.

*Summary:* Our findings show that devising hybrid methods helps practitioners reach set goals. However, even the best strategies applied today are still suboptimal and are not guaranteed to reach these goals. Furthermore, deviation happens also for hybrid processes, yet, was observed to be counterproductive in terms of achieving the goals.

*Lessons Learned for Practitioners:* When devising hybrid methods, it seems to be the best strategy to *first plan the hybrid method and then evolve it based on experience*. Whenever possible, it seems that it is better to *mitigate process deviation*. If deviation happens, it is worthwhile to investigate the reasons. For example, if developers perceive the standard process too complex, it might help to revise the process *together* and to plan an adaptation.

### B. Limitations and Threats to Validity

We discuss threats to validity of this study following the structure proposed by Wohlin et al. [23].

*Construct Validity:* The general threat to construct validity of questionnaire-based research is the risk that questions are misunderstood by the participants. To mitigate this risk, we designed the questionnaire with a team of multiple researchers, involving internal and external reviews. We performed pre- tests as described in Section 3 and, afterwards, conducted a first survey with 69 subjects, which led to a revision of the questionnaire. Potential misunderstandings due to language issues were addressed by providing the questionnaire in four languages (translated by native speakers). Due to these mitigation strategies, we are confident that risks are mitigated.

Another risk is that the participants do not reflect the desired target population, since the link for participation was spread via multiple networks and mailing lists. Thus, the survey could have been answered by persons out of our target population potentially introducing biases to the results. Based on the specific knowledge required to answer the questionnaire and the consistently meaningful free-text answers, we consider this threat very small and, thus, mitigated.

*Internal Validity:* The preparation of the data and data-cleaning can be considered a threat to the internal validity as errors might have been introduced. Furthermore, the choice and application of statistical tests can introduce errors. To avoid this threat, all steps of the analysis have been performed by two or more researchers and were later reviewed by further researchers. Due to these review processes, we have confidence that the method is reliable and reproducible. A risk to the qualitative analysis could be incomplete assessments of relevant themes and the incorrect summary of observations. To minimize both, the qualitative analysis was conducted with multiple researchers performing two rounds of coding. Data was coded by multiple researchers. Finally, the summary of the data was performed by a team of three researchers.

*External Validity:* The generalization of a single study to all cases of software development is always a threat. However, we reached a broad coverage of domains and participant roles as well as an even distribution of company sizes. This allows for making observations that are independent of these factors. Concerning the generalizability of results across countries it would have been interesting to have more data points from Africa, Asia, and North America. Having few data points from countries in these regions threatens the global generalizability of our results. However, the data points that we have, e.g., from Uganda, indicate that our results might be to some degree valid for these regions. Future studies are needed to confirm this.

*Conclusion Validity:* For the statistical tests we worked with a significance level of $p \leq 0.05$. The identified significant results will have to be confirmed in future studies. However, non-negligible effect sizes were observed, indicating that the results are potentially relevant. The choice of the thresholds can of course be discussed. Nonetheless, we contend that the effects observed provide a baseline for future studies.

### C. A Baseline for Future Research

Since our results show that hybrid methods are the current state of practice, we suggest taking these findings as a new baseline for future research on hybrid, flexible and adaptive software development processes. In the following, we discuss arising research challenges.

The strategies applied today are still some way from perfect when it comes to devising hybrid methods. Therefore, the first main future direction for research is to *provide better strategies to devise hybrid methods*.

Studying the reasons for process deviation well help better understand potential directions to mitigate the use of strategies for process deviation. Among the most interesting observations in our data is that goals for deviation are not necessarily different from goals for the use of a hybrid method. Deviation seems to be used where also a planned evolution could be appropriate. Similarly, deviation to accommodate context factors or to improve the process could—in theory—also happen in a more planned way as guidelines and even standards for such initiatives are in place. It would be interesting to further *investigate why such a planned evolution is not happening instead of the deviation*.

Some cases of deviation are caused by external triggers and requests that appear while the project is running. As a research community, we need to *help companies to develop strategies to deal with such situations in a better, more efficient and effective way*, e.g., by helping practitioners document deviation experiences from one project and providing better planned alternative solutions for future projects that might become subject to similar emergencies.

If deviation is frequently happen in a company's projects, it is worth reconsidering the standard process, as it might not provide enough guidance or could be too complex. Processes that aim to cover many (very) different project settings have to be considered prone to deviation. Research should *provide straightforward guidelines or metrics to help practitioners identify processes at risk for deviation*, based on companies' projects and structures.

Finally, we should research and develop strategies suitable for use by practitioners when being confronted with the need to *integrate different processes or organizations*.

## 6. CONCLUSION

Companies usually do not apply development approaches by the book. In fact, they often combine different development frameworks, methods and practices. Among the most frequently used frameworks and methods are Scrum and Iterative Development. However, they are mostly combined with other frameworks, methods and practices into so-called *hybrid development methods*. While the chosen hybrid methods are generally suitable to reach the company's goals with a probability of 66%, project/product teams not deviating from standard policies seem to be a bit more successful in achieving their goals. The reasons for deviating from a process are, among others, explicit goals, context factors as well as issues with the standard process. However, the goals companies strive for do not depend on the deviation from policies.

In a nutshell, devising hybrid methods helps practitioners reach their goals. However, even the best strategies for devising hybrid methods are imperfect. Consequently, our findings pose a new baseline for further research, which is necessary to identify the best practices for devising hybrid development methods.

## ACKNOWLEDGMENTS

We thank all the study participants and the researchers involved in the HELENA project for their great effort in collecting data points. *Dietmar Pfahl* was supported by the group grant IUT20-55 of the Estonian Research Council. *Rafael Prikladnicki* is partially funded by FAPERGS (17/2551-0001/205- 4) and CNPq. *Joyce Nakatumba-Nabende* was supported by the Sida/BRIGHT project 317 under the Makerere-Swedish bilateral research programme 2015-2020. *Fergal McCaffery* and *Özden Özcan Top* were supported by Science Foundation Ireland grant 13/RC/2094 and co-funded under the European Regional Development Fund through Lero, the Irish Software Research Centre.